\documentclass[
superscriptaddress,
 amsmath,amssymb,
 aps,
prx,
notitlepage
]{revtex4-1}

\usepackage{graphicx}
\usepackage{dcolumn}
\usepackage{bm}
\usepackage[mathlines]{lineno}
\usepackage{color}
\usepackage{soul}
\usepackage{enumitem}

\begin{document}


\title{A Heuristic Quantum-Classical Algorithm for Modeling Substitutionally Disordered Binary Crystalline Materials}

\author{Tanvi P. Gujarati}%
 \affiliation{IBM Research Almaden, San Jose, California 95120, USA}%
\author{Tyler Takeshita}%
 \affiliation{Mercedes-Benz Research and Development North America, Sunnyvale, California 94085, USA}
\author{Andreas Hintennach}%
 \affiliation{Daimler AG, HPC G012, 45 Hanns-Klemm-Str., Boblingen 71034, Germany}
\author{Eunseok Lee}%
 \email{eunseok.lee@daimler.com}
 \affiliation{Mercedes-Benz Research and Development North America, Sunnyvale, California 94085, USA}
 \affiliation{Department of Mechanical and Aerospace Engineering, The University of Alabama in Huntsville, Huntsville, Alabama 35899}

\date{\today}

\begin{abstract}
Improving the efficiency and accuracy of energy calculations has been of significant and continued interest in the area of materials informatics, a field that applies machine learning techniques to computational materials data. Here, we present a heuristic quantum-classical algorithm to efficiently model and predict the energies of substitutionally disordered binary crystalline materials. Specifically, a quantum circuit that scales linearly in the number of lattice sites is designed and trained to predict the energies of quantum chemical simulations in an exponentially-scaling feature space. This circuit is trained by classical supervised-learning using data obtained from classically-computed quantum chemical simulations. As a part of the training process, we introduce a sub-routine that is able to detect and rectify anomalies in the input data. The algorithm is demonstrated on the complex layer-structured of Li-cobaltate system, a widely-used Li-ion battery cathode material component. Our results shows that the proposed quantum circuit model presents a suitable choice for modelling the energies obtained from such quantum mechanical systems. Furthermore, analysis of the anomalous data provides important insights into the thermodynamic properties of the systems studied.

\end{abstract}

\maketitle


\section{\label{sec:intro}Introduction}
Materials Informatics (MI)~\cite{Ward2017,Gomez-Bombarelli2018,Pilania2013}, a recent trend in computational materials science, is a field of study that applies machine learning techniques to materials data in order to efficiently predict material properties~\cite{Jain2013,Kirklin2015,Ramprasad2017,PhysRevB.99.064103,Schutt2017}, ultimately aimed at accelerating materials development and deployment, e.g. for battery materials~\cite{doi:10.1002/eem2.12053,doi:10.1002/aenm.201200593}. One of the challenging tasks in MI is to construct sufficiently rich models that can describe material properties at atomic scales which are generally calculated using quantum mechanical principles. Often, this requires expensive computations and data management that are intractable with today's computational resources~\cite{Xue2016,Hill2016,Lookman2017,Takahashi}. 

A quantum computer may be an attractive candidate to mitigate some these challenges by providing access to quantum learning models which are more suitable for studying data generated from quantum mechanical processes and models which are potentially inaccessible for the classical computational techniques in terms of computational prowess. Although the ability to implement pragmatic problems on current quantum computers is limited due to decoherence and errors~\cite{Temme2017,Colless2018,Li2017}, recent approaches~\cite{Farhi2014,McClean2016, PhysRevA.98.032309}  based on the hybridization of quantum and classical computation have been successfully demonstrated on the quantum hardware currently in operation~\cite{Kandala2017,Havlicek2019,Peruzzo2014, Kandala2019, Gao2018, Preskill2018, Arute2019, Pednault2019}. In these hybrid approaches, a single lengthy operation on a quantum processing unit is replaced with a series of shorter processes which are less vulnerable to decoherence and errors, interleaved with classically computed optimization routines.

Aligned with these approaches, we present here a hybrid quantum-classical algorithm to model the energy of substitutionally disordered binary crystalline materials. More specifically, a quantum circuit that links the energy and atomic position of materials is designed and trained by supervised-learning based on classically computed data instances (results of quantum chemical simulations from classical computations). Quantum chemical simulations on classical computers are often very expensive and verifying the accuracy of the obtained results is very difficult. By training an efficient quantum circuit model that can effectively learn the properties of interest for a given chemical system we aim to address these two issues. The immediate advantage of using a quantum circuit model is the availability of a quantum mechanically enhanced feature space for data representation. This enhanced feature space can exploit correlations generated due to quantum entanglement which a classical model is unable to access. In the algorithm described below, data generated from the classical computations can be fed directly into the quantum circuit model and hence problems associated with data input and output prevalent in many of the quantum machine learning algorithms does not impact this implementation \cite{Biamonte2017}. To demonstrate the developed algorithm we apply it to the layered  Li$_x$Co$_{2-x}$O$_2$ (LCO) system, an important material used in Li-ion batteries.

This algorithm is explored and explained with an emphasis on two aspects: (i) quantum circuit model design and scaling of the number of parameters and (ii) anomaly detection in the input data instances. The significance of anomaly detection is discussed, particularly, in relation to the magnetic moments of Co ions in the LCO system.

\section{Background}
The energy of single-crystal materials can be expressed as a function of occupation variables $\sigma_j$, which indicate an atomic species on lattice site $j$. The total number of distinct configurations $\sigma=\sigma_1\sigma_2\cdots\sigma_N$ for a lattice with $N$ lattice sites is equal to $m^N$, where $m$ is the number of different atomic species constituting the material that can occupy a given site. Thus, to completely describe the energy of the system, $E(\sigma)$, $m^N$ parameters will be required. For example, a binary crystalline material with its atoms dispersed on $N$ lattice sites will have $2^N$ distinct $\sigma$ configurations that can be distinguished by $2^N$ parameters. If the force-field of the system is already known to have an explicit formula as a function of continuous variables, e.g. pairwise potential as a function of interatomic distance, fewer number of parameters may be sufficient for modeling the energy. However, without such foreknowledge, in principle, $m^N$ parameters are needed. In practice, the number of the required parameters is reduced by using intrinsic material properties, spatial symmetries and data-science techniques. One such technique is the cluster expansion method~\cite{Sanchez1979,Wolverton1994,Asta2001,VanderVen2010,Lee2012a} where the energy is expanded as a linear combination of orthogonal functions (referred to cluster functions) designed to represent atomic clusters, with corresponding expansion coefficients used as parameters to model the energy. The total number of atomic clusters that can be created is equal to $m^N$ and correspondingly there are $m^N$ cluster functions (that is, $m^N$ parameters). To reduce the number of parameters, the set of cluster functions is truncated by considering rotational and translational symmetries of the atomic clusters, assuming a cutoff distance for effective interactions between atoms, and iteratively searching for the most representative atomic clusters based on a trade-off between efficiency and accuracy of the model. Although these techniques have been demonstrated successfully in certain areas of materials research, in particularly in the area of battery materials ~\cite{Asta2001,Lee2012a,Persson2010,Lee2017}, there are still fundamental limits in increasing computational efficiency further, due to the time-consuming process of sorting out the most representative cluster functions as well as the risk of propagating errors caused by incorrectly truncated set of cluster functions~\cite{10.1145/1541880.1541882,chalapathy2019deep}. Recently, techniques going beyond traditional cluster expansion that model the energies using artificial neural networks and machine learning techniques have been explored~\cite{Natarajan2018,acs.jpcc.9b03370,Lee_2020}. While these approaches provide a way to correlate features systemically, prior selection of input features or prior truncation of features in convolution operations still remains necessary. In our approach presented below, the design of the quantum circuit model is implemented independently of the prior selection or truncation of features (cluster functions). 

In recent years quantum circuits that encode single qubit rotations and two qubit interactions with free parameters have been widely used for determining the lowest energy state of chemical and physical systems using algorithms like Variational Quantum Eigensolver (VQE) and Quantum Approximate Optimization Algorithm (QAOA)\cite{Kandala2017,Peruzzo2014,Barkoutsous2018,Farhi2014,Zhou2020}. These quantum circuit models with free parameters are also ubiquitous in the quantum optimization and quantum machine learning literature for classification and regression problems \cite{Havlicek2019, PhysRevA.98.032309, Zoufal2019, Perez-Salinas2020,Anschuetz2019, Benedetti2019}. Quantum circuit families can be carefully chosen such that they provide the freedom to explore a specific subset of states in the Hilbert-space using correlations that cannot be described classically, can be customized to account for the constraints set by error prone near-term quantum hardwares and can also guarantee that simulating these quantum circuits with classical computers scales inefficiently \cite{Havlicek2019,Bravyi308},  hence, providing a path to quantum advantage. In this work, we choose a quantum circuit model that is based on spin Hamiltonians, can be efficiently executed on near-term quantum hardware and is inspired by circuit families that are known to be hard to simulate on classical computers.


\section{\label{sec:approach}Approach}
\begin{figure*}
    \centering
    \includegraphics[width=4in]{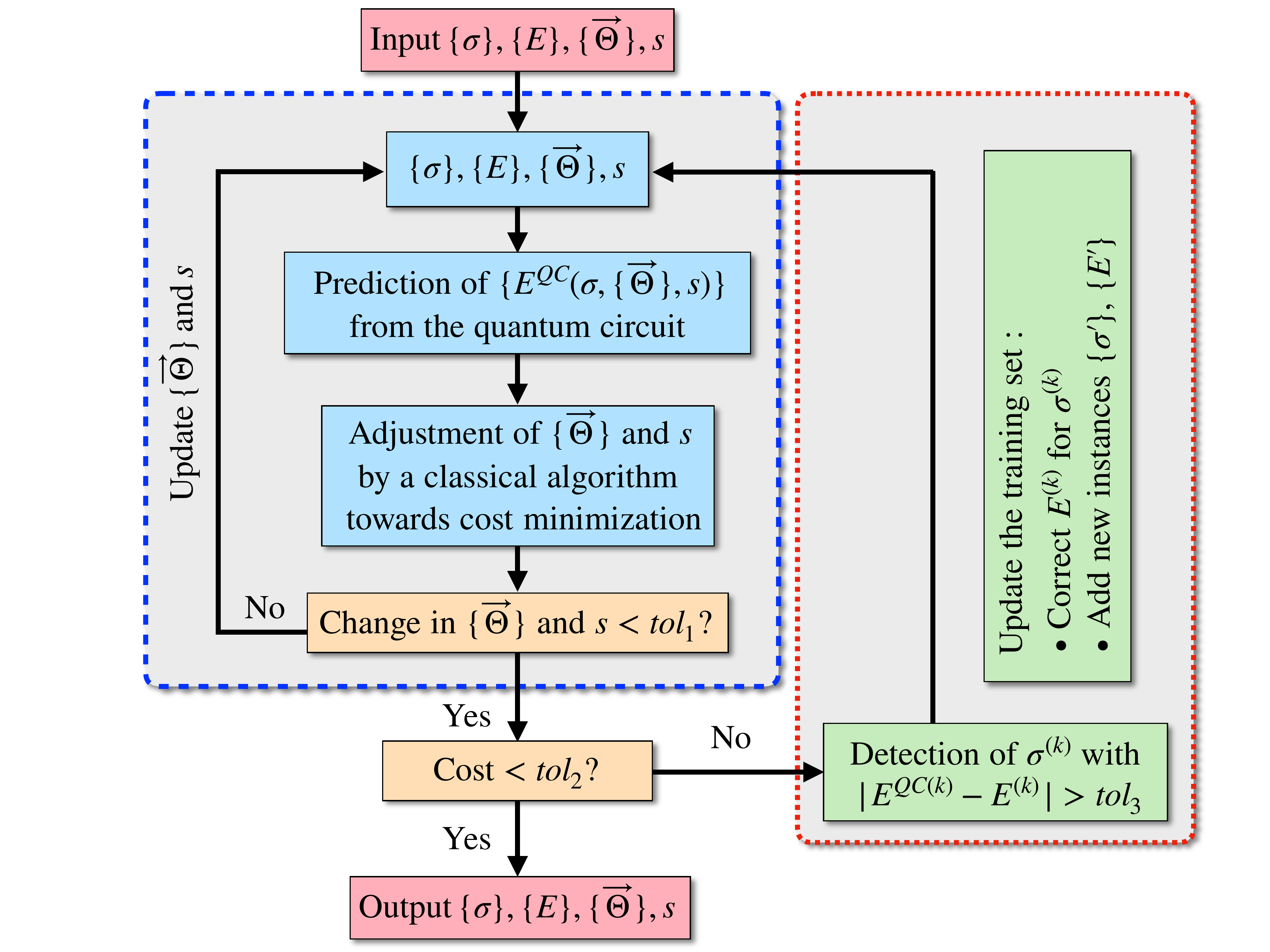}
    \caption{Workflow of the proposed algorithm - $\sigma$ and $E$ are the configuration of occupation variables and the energy from classically-computed quantum chemical simulation (density functional theory calculation in this study), respectively, per each data instance. \{$\sigma$\} and \{$E$\} constitutes the training set. The parameters $\{\vec{\Theta}\}$ and $s$, which are the coefficients of the quantum circuit and a scaling factor, respectively, are used to predict the energy $E^{QC}$ for a given $\sigma$ and optimized classically through minimization of a cost function. $tol_1$ and $tol_2$ are the thresholds to determine if the parameters were converged and the cost reached the minimum, respectively, while $tol_3$ is used to identify data instances with anomaly. $E^{QC(k)}$ and $E^{(k)}$ are the energies for the $k$-th configuration $\sigma^{(k)}$ in the training set. The training set is updated by correcting the anomalous data instances and adding new data instances. Variational quantum-classical optimization process and data anomaly detection process are denoted by blue-dashed and red-dotted regions, respectively.}
    \label{fig:workflow}
\end{figure*}

We develop a hybrid quantum-classical algorithm to efficiently model the energy of a binary crystalline material as a function of the configuration, $\sigma~(\sigma=\sigma_1\sigma_2\cdots\sigma_N,~ \sigma_j \in \{+1,-1\})$. Values of +1 or -1 are assigned based on the atomic species at a given lattice site. We assume that each lattice site is occupied by either of the two species and no lattice site remains empty.  A parameterized quantum circuit is designed as a function of the configuration and trained via minimization of a cost function that measures the difference between the energies predicted by the quantum circuit and those computed by classical techniques for a training set. Once converged, the trained model will be able to predict the energy of arbitrary $\sigma$ without performing classical energy calculations. 
%
Workflow of the proposed algorithm is as illustrated in FIG.~\ref{fig:workflow}. 
%
%

Our design principle for quantum circuit assigns one qubit to each occupation variable and entangles every pair of the nearest neighboring qubits. The designed quantum circuit model consists of two consecutive layers of parameterized single and two qubit operations acting on $N$ qubits where each layer is given by the following unitary transformation:
%
%
%
\begin{widetext}
\begin{equation}
    {U(\{\vec{\Theta}\},\sigma)}={\exp\left(i\sum_{j=0}^{N-1}\phi_{j}\hat{X}_{j}\right)\exp\left(i\sum_{j=0}^{N-1}\theta_j\sigma_{j}\hat{Z}_{j} + i\sum_{j=0}^{N-2}\theta_{j,j+1}\sigma_{j}\sigma_{j+1}\hat{Z}_{j}\hat{Z}_{j+1}\right)}
    \label{eq:onelayer}
\end{equation}
\end{widetext}
In Eq.~(\ref{eq:onelayer}), $\sigma_j$ denotes the occupation variable on lattice site $j$, and $\hat{X}_{j}$ and $\hat{Z}_{j}$ denote the Pauli operators at the lattice site $j$. $\{\vec{\Theta}\}=\{\phi_j,\theta_j,\theta_{j,j+1}\}$ indicate the free parameters in the circuit to be learned. The terms corresponding to the Pauli operators $\hat{X}_{j}$ and $\hat{Z}_{j}$ are represented by single qubit rotations around the $x$ and the $z$ axis of the $j^{th}$ qubit, respectively, in the quantum circuit. The rotation angles are controlled by the parameters $\phi_{j}$, $\theta_{j}$ and occupation variable $\sigma_{j}$. The term with $\hat{Z}_{j}\hat{Z}_{j+1}$ interaction between the nearest neighboring qubits is encoded with a single $Z$ rotation gate and a rotation angle of $\theta_{j,j+1}\sigma_{j}\sigma_{j+1}$ on the qubit representing lattice site $j+1$ sandwiched between two CNOT gates with qubit representing the $j^{th}$ lattice site as the control qubit and the qubit representing the $j+1^{th}$ lattice site as the target qubit (refer to the example circuit in Fig. (\ref{fig:qc_exp_detail}) shown in Appendix \ref{sec:qc_circuit_properties}). We use two consecutive layers of the unitary circuit presented in Eq.~(\ref{eq:onelayer}) to represent quantum chemical interactions in the lattice system. Therefore, there are overall $6N-2$ free parameters in this circuit for a lattice system with $N$ lattice sites. While there can be several variations in the arrangement of the entangling blocks in this circuit model, we adopted one in which the circuit depth increases linearly with the number of qubits, for simplicity.  Based on the results in~\cite{Havlicek2019,Bravyi308,Bravyi2019}, we believe that this circuit model would be hard to simulate classically as the system size grows. The suggested quantum circuit model is heuristically chosen, but its structure with interleaving unitaries as given in Eq. (\ref{eq:onelayer}) is inspired by the philosophy of QAOA where unitaries describing the Hamiltonian of the system are interleaved with a simple mixing Hamiltonian \cite{Farhi2014}. The assignment of lattice sites to qubits can be chosen based on the crystal structure of the system of interest. Refer to Appendix~\ref{sec:qc_circuit_properties} for more discussion on the properties of the quantum circuit used.
%
%
%
The overall structure of the designed quantum circuit is as illustrated in FIG.~\ref{fig:quantumcircuit} and detailed circuit for a 4 lattice site system is provided in Appendix~\ref{sec:qc_circuit_properties}.
\begin{figure*}
    \centering
    \includegraphics[width=7.05in]{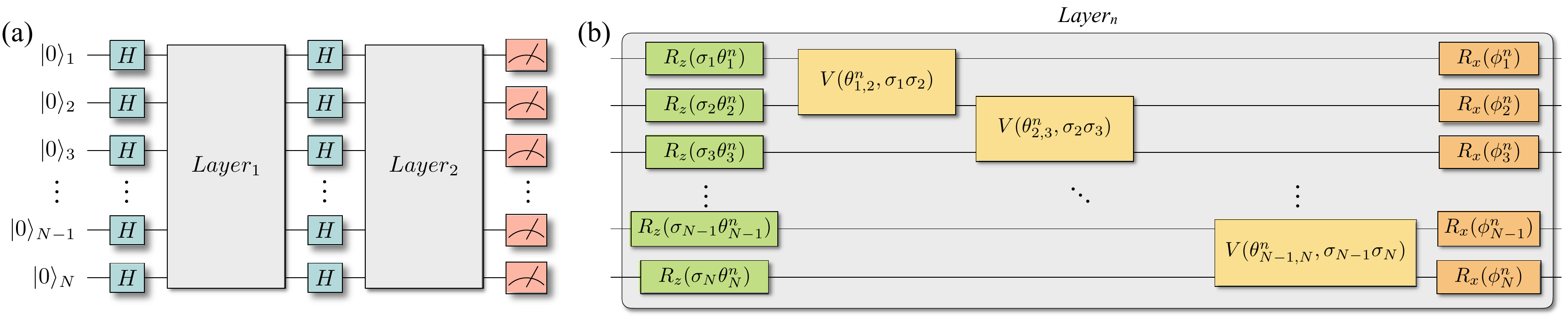}
    \caption{Illustration of the proposed quantum circuit - (a) the entire circuit and (b) one layer. The circuits consists of a set of free parameters $\{\vec{\Theta}^n\}=\{\theta_{j}^{n},\theta_{j,j+1}^{n},\phi_{j}^{n}\}$ where the subscript denotes the lattice site(s) and the superscript specifies the layer number. The two-qubit operator $V(\theta_{j,j+1}^{n},\sigma_j\sigma_{j+1})$ denotes $e^{i\theta_{j,j+1}^{n}\sigma_{j}\sigma_{j+1}\hat{Z}_j\hat{Z}_{j+1}}$ and represents the entangling operator between qubits $j$ and $j+1$ on the layer $n$.}
    \label{fig:quantumcircuit}
\end{figure*}
The energy of the lattice system for a given configuration $\sigma$ is evaluated as the expectation value of the state $|\psi(\sigma)\rangle$ given in Eq. (\ref{eq:state})
\begin{equation}
    |\psi(\sigma,\{\vec{\Theta}^{n=1,2}\})\rangle = U(\{\vec{\Theta}^{2}\},\sigma)H^{\otimes N}U(\{\vec{\Theta}^{1}\},\sigma)H^{\otimes N}|0\rangle^{\otimes N}
    \label{eq:state}
\end{equation}
with respect to the $\hat{X}_{1}\hat{Y}_{2} \cdots \hat{X}_{N-1}\hat{Y}_{N}$ operator using a quantum processing unit. 
The operator $\hat{X}_{1}\hat{Y}_{2} \cdots \hat{X}_{N-1}\hat{Y}_{N}$ was chosen to reduce the effects of the unintentional parity-symmetries of occupation variables and hence to distinguish different $\sigma$ effectively (see Appendix~\ref{sec:qc_circuit_properties}). If a priori knowledge about the Hamiltonian describing the chemical system is available, it can be incorporated into selection of the measurement operators. The energy predicted by this model, $E^{QC}$, is given as: 
%
\begin{widetext}
\begin{eqnarray}
{E^{QC}(\sigma, \{\vec{\Theta}^{n=1,2}\},s)} &= & s\langle \psi(\sigma, \{\vec{\Theta}^{n=1,2}\})|\hat{X}_{1}\hat{Y}_{2} \cdots \hat{X}_{N-1}\hat{Y}_{N}| \psi(\sigma, \{\vec{\Theta}^{n=1,2}\})\rangle
\label{eq:e_qc}
\end{eqnarray}
\end{widetext}
The exact functional form of the obtained energy is beyond the scope of the current work. In Eq.~(\ref{eq:e_qc}), we introduce a scaling factor, $s$, which is also an optimization parameter, to account for the overall scale of the predicted energies since each material system would have its corresponding energy scale. With the inclusion of the scaling factor, the total number of parameters to be optimized is $6N-1$. $E^{QC}$, as a function of $\sigma$, is then compared with the energy obtained from classically-computed quantum chemical simulations. The quantum circuit part of this algorithm was implemented using IBM's open source Qiskit Aqua software and the results were simulated using the Statevector simulator provided within Qiskit \cite{Qiskit}. In this study, density functional theory (DFT) was used to calculate the energy classically, $E^{DFT}$. To help with the optimization procedure, the classical DFT energy data was pre-processed. Details of this procedure are outlined in Appendix~\ref{sec:DFT_calculations}.

Data instances from the DFT calculations were divided into two sets, the training set and the test set. Both $E^{QC}$ and $E^{DFT}$ were obtained for all configurations in the training set. The coefficients ($\{\vec{\Theta}^{1}\},\{\vec{\Theta}^{2}\}$) and the scaling parameter $s$ were then optimized to minimize the cost, defined as the Root-Mean-Square-Error (RMSE) of $E^{QC}(\sigma)$ compared with the $E^{DFT}(\sigma)$ for each configuration $\sigma$ in the training set, as follows.
\begin{equation}
    cost=\sqrt{\sum_{i=1}^{N_{data}}\frac{(E^{QC(i)}-E^{DFT(i)})^2}{N_{data}}}
    \label{eq:cost_function}
\end{equation}
In Eq.~(\ref{eq:cost_function}), $N_{data}$ is the number of data instances in the training set and $i$ denotes the $i^{th}$ data instance. This optimization process is carried out using a combination of Constrained Optimization BY Linear Approximation (COBYLA)~\cite{Conn1997} and Adaptive Moment Estimation (Adam)~\cite{Kingma2015}.  
%
%
The COBYLA optimization algorithm is applied until the number of iterations for parameter optimization reach the preset maximum of 10,000, or satisfies a preset convergence error tolerance, $tol_1$ = $10^{-4}$. In the former case, the optimization parameters are then refined further using Adam algorithm until convergence set by $tol_1$ is reached.
After the completion of the optimization run (refer to FIG.~\ref{fig:workflow}), if the evaluated cost is found to be higher than a fixed value, $tol_2$, all data instances used for the training which however show a discrepancy of greater than $tol_3$ between $E^{QC}$ and $E^{DFT}$ are examined by performing follow-up DFT calculations for the same $\sigma$ but with updated DFT parameters, particularly, the initial values of magnetic moments in this study. If the $E^{DFT}$ from the follow-up DFT calculation is lower than the current value, the data point is replaced with new value in the training set. Otherwise, a new data instance is created by performing a DFT calculation for a similar $\sigma$ and added to the training set. We refer to this process as anomaly detection and treatment (see Appendix~\ref{sec:anomaly_treatment} for more detail). The sequence of parameter optimization followed by anomaly detection and treatment (referred to as a round) is repeated until the obtained cost is less than a preset value $tol_2=0.03$ eV per cation (or 0.015 eV per atom). This value for $tol_2$ was adopted to maintain the ratio of $tol_2$ to the variation in the $E^{DFT}$ values as low as 1.5$\%$ and thus ensure that the circuit model can effectively distinguish the energies of distinct $\sigma$. $E^{DFT}$ values in the training set tended to vary by ~2 eV per cation. The tolerance $tol_2$ can be set to a lower value at the price of slower convergence. $tol_3$ was set to be flexible during rounds. In initial rounds, a relatively bigger value was used because the model, which is used as the reference, would be still far from convergence. When the model reached closer to convergence in the later rounds, a smaller value was used. In this study, $tol_3$ was set to 0.1 eV per cation initially and reset to 0.06 eV per cation after a significant reduction of the cost was observed at the third round, to detect anomalous data more rigorously.


The algorithm and quantum circuit are applied to the LCO system, a core component in one of the widely used Li-ion battery cathodes, lithium Nickel-Manganese-Cobalt oxides abbreviated as NMC. While the LCO system has been modeled in many previous computational research studies, its magnetic properties have rarely been studied due to difficulties in calculating magnetic moments precisely (refer to Appendix{~\ref{sec:DFT_calculations}}). Anomaly detection and treatment process in our algorithm is expected to mediate these difficulties. Ideally, the layered LiCoO$_2$ (LCO at $x=1$) consists of alternating cation (Li or Co) layers with one anion (O) layer between each cation layer. However, in our model cationic lattice sites are assumed to be occupied by either Li or Co considering cations-intermixing, in particular, when the chemical composition of LCO deviates from $x=1$. Oxygen atoms are assumed to reside in anion layers and, thus, only cationic lattice sites are assigned $\sigma_j$. While $\sigma_j$ can have any value, except 0 as it nullifies the associated coefficients in the quantum circuit, we adopted +1 for Co and -1 for Li in this study, for convenience.

\section{\label{sec:result}Results and Discussion}
%
We applied the developed algorithm to a 4 and 8 cationic lattice site LCO system.  
In the 4 lattice site system, the total number of possible configurations, $\{\sigma\}$, is 16. Therefore, in principle, 16 parameters should be sufficient
to model the energy of the system. However, the spin-polarized DFT calculations (used in this study) for transition-metal oxides often fail to converge to the ground state due to the slow convergence of magnetic moment (see Refs~\cite{Zeller2006,Wills2010,Lee2017,Bihlmayer2018}) and predict a number of different thermodynamically-meta-stable states for identical $\sigma$. The added complexity of different magnetic states produces many more than $2^N$ data instances for $N$ lattice site system, including anomalous data instances due to unconverged magnetic moments. Thus, the anomaly detection and treatment process is essential for reliable modeling of the energies. In a parallel effort, we also modeled this 4 lattice site system classically (16 parameters was small enough for classical approaches) to provide a benchmark result for our algorithm. The parameters and expansion basis were formulated using the cluster expansion method but without any truncation, and the same anomaly detection and treatment as our algorithm was processed. We would like to highlight that the anomaly detection and treatment sub-routine is applicable to classical algorithms as well, in general. One distinction being, with traditional approaches such as truncated cluster expansion, the anomaly detection would be processed with a model that is intrinsically incomplete and thus possesses a high risk of bias.

%

The training set used consisted of 30 data instances. 
%
%
Note that a few instances of the training set had identical $\sigma$ but different energies due to distinct magnetic moments of ions, exposing the need for anomaly detection. After the parameter optimization in the 1st round, the value of the cost function reduced to 0.10 eV/cation from the initial value of 0.38 eV/cation.
We also observed that data instances that displayed a large discrepancy between $E^{DFT}$ and $E^{QC}$ were often not fully-converged in their magnetic moments. For example, a large $\left|E^{QC}-E^{DFT}\right|$ was observed for LiCoO$_2$ (LCO at $x=1$) data instances with magnetic Co ions, while the oxidation state of Co ions at $x=1$ is known to be Co$^{3+}$ which is nonmagnetic. 
Those data instances with large $\left|E^{QC}-E^{DFT}\right|$ were updated by the procedure for anomaly detection and treatment. This sequence was repeated to the fourth round and the resultant cost reduced to 0.018 eV/cation (or 0.009 eV/atom), corresponding to 1.55\% of the variation of $E^{DFT}$ in the training set, which ranged from -11.596 to -10.431 eV per cation. The corresponding mean absolute percentage error (MAPE) was 0.096\%. 
As stated earlier, we also modeled $E(\sigma)$ classically. The resultant cost was 0.12 eV/cation after the first round and then reduced to 0.018 eV/cation after the fourth round, which are almost the same as the result from our algorithm.

\begin{figure*}
    \centering
    \includegraphics[width=5in]{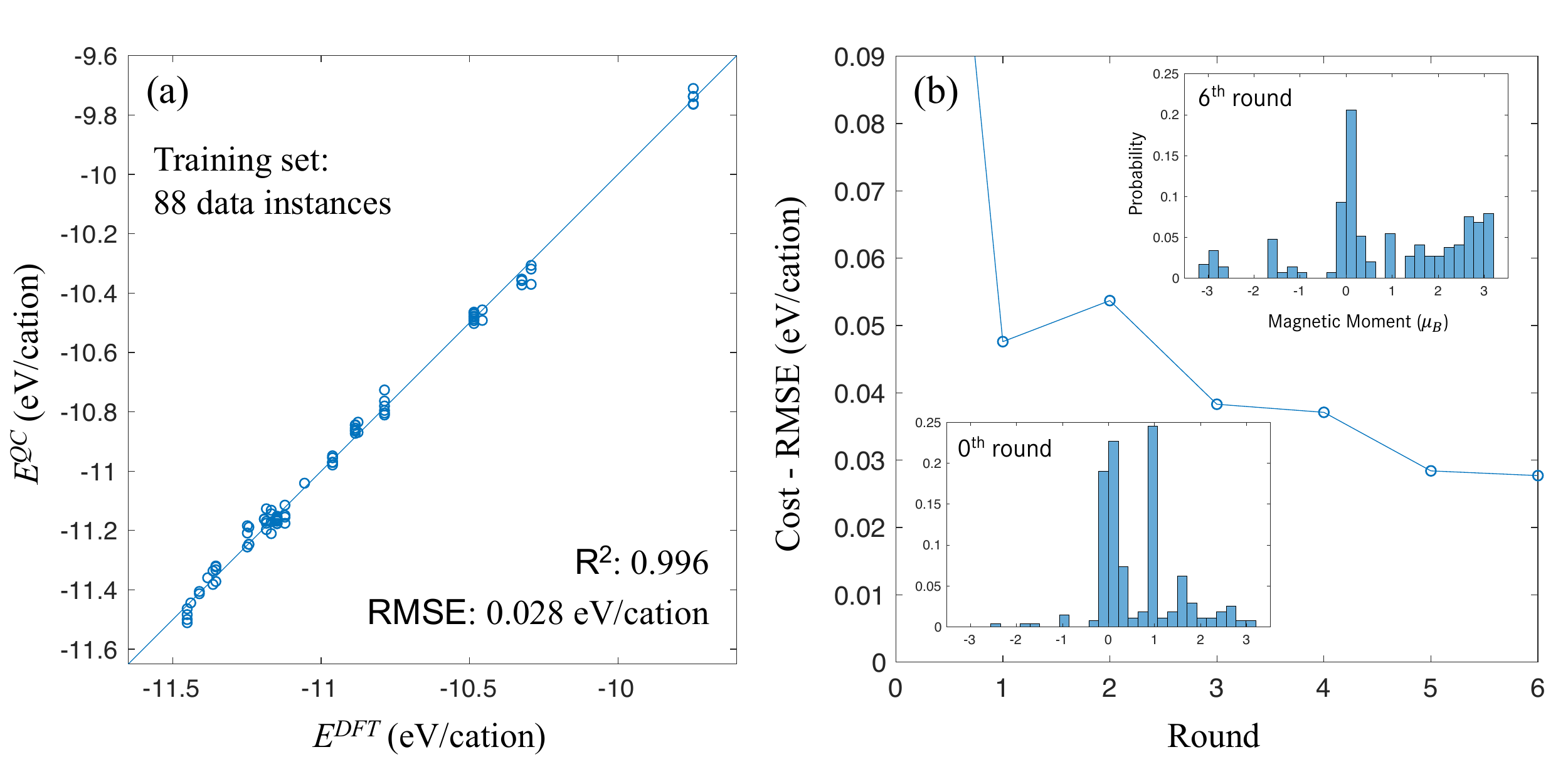}
    \caption{(a) Comparison between $E_i^{DFT}$ and $E_i^{QC}$ for data instances in the training set for 8 cationic lattice site LCO system. RMSE and R$^2$ denote the resultant cost and the coefficient of determination, respectively. (b) The resultant cost after each round. The value at 0th round was adopted from the cost at the 1st step of the optimization of the parameters which started from random values. Two inset figures show the probability histogram of the magnetic moments of Co ions in the training set, obtained from DFT calculations using GGA+U scheme, at the 0th and the 6th rounds.}
    \label{fig:training_set_result}
\end{figure*}

The algorithm was then applied to a 8 cationic lattice site LCO system. As stated earlier, many more than 2$^8$ data instances can be produced from DFT calculations due to the slow convergence of magnetic moment. The initial training set had 72 data instances of randomly chosen and distinct $\sigma$. The anomaly detection and treatment process was then applied.
%
%
After 6 rounds tolerance criterion for the cost function was met. The final size of the training set increased to 88 data instances and the resultant cost was 0.028 eV/cation. The performance of this training result can be assessed by comparing $E^{DFT(i)}$ and $E^{QC(i)}$, which is shown in FIG.~\ref{fig:training_set_result}(a).
The cost 0.028 eV per cation corresponds to 1.58\% of the variation in $E^{DFT}$ values of the training set that ranged from -11.451 eV to -9.746 eV per cation. The corresponding MAPE was 0.20\%.

We remark that in our algorithm the energy of the system is not explicitly decomposed into the parameters of interatomic interaction energies which is typically in classical approaches like cluster expansion; where they are pre-truncated before fitting or selected via an iterative search process. Instead, our algorithm represents the energy of material system by a parameterized quantum circuit as a whole, avoiding the risk of severe pre-truncation of parameters and time-consuming iterative selection process. Modeling a system as a whole also brings computational efficiency in anomaly detection and treatment by reducing the risk of over-fitting to a specific set of pre-selected features or functions.

At the end of each round, we investigated if there was a relationship between the anomalous data and the chemical composition of materials.
In general, $\left|E^{QC}-E^{DFT}\right|$ tended to be larger in the Co-rich region ($2-x\geq1.5$). This tendency is compatible with the fact that CoO, which is the LCO system with the highest Co content, has a cubic structure, clearly different from the hexagonal plane structure of the host LCO system. On the other hand, Li$_{2}$O$_{2}$, the LCO system with the highest Li content, also has a different crystal structure from the LCO but still contains partial geometric similarities to the host LCO system, such as hexagonal planar structure and octahedral cationic sites in every other cation layers. Thus the geometric incompatibility with the host LCO system may be less significant in the Li-rich region. An illustration of their geometric structures is provided in Fig. (\ref{fig:geo_structures_three}). As explained in Appendix~\ref{sec:DFT_calculations}, in the training, the energies of geometrically optimized structures are used. Thus, as the rounds proceed, the coefficients are trained to represent the energies of the structures which are more geometrically-compatible with the host LCO system. The result also infers that data instances can be grouped into roughly two regions, the Co-rich region and the other region, based on the geometric compatibility with host LCO system. Note that this grouping of data instances was learned through the rounds and not by a presumption on the geometric structure. This result can be considered as another evidence for the feasibility of the developed algorithm.
\begin{figure*}
    \centering
    \includegraphics[width=6in]{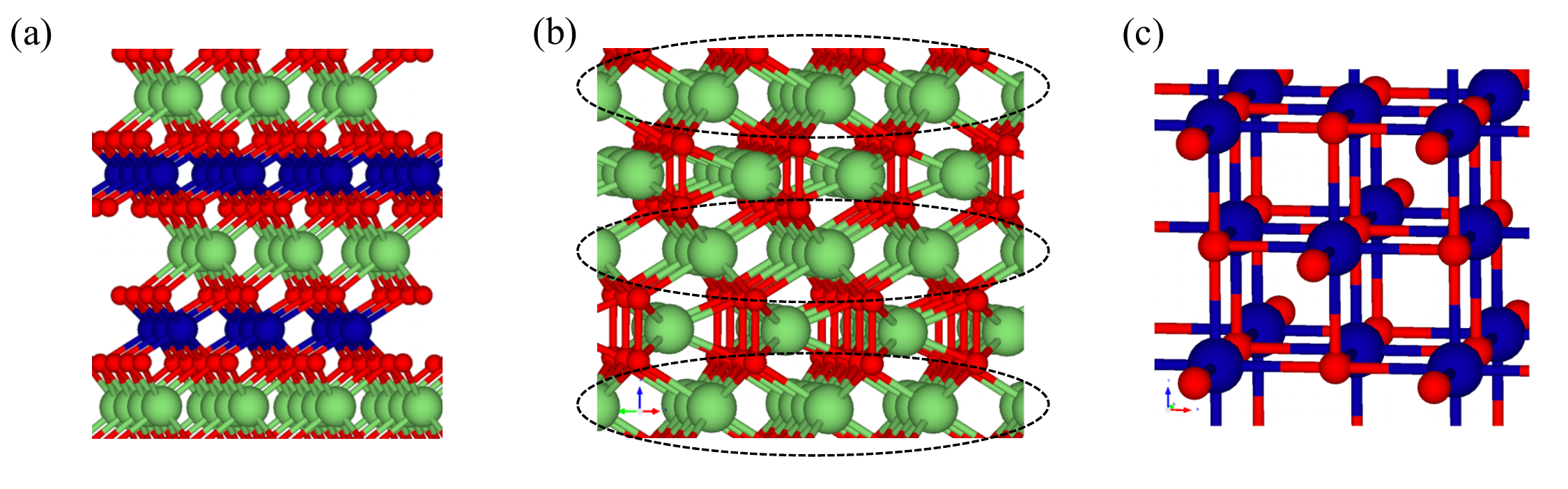}
    \caption{Illustration of the crystal structures of the LCO systems used in the training - (a) the ideal LiCoO$_2$, (b) Li$_2$O$_2$, and (c) CoO. In (b), the dashed oval regions are to indicate LiO$_6$ or CoO$_6$ octahedra arranged toward a hexagonal cation distribution on the basal plane, a partial geometric-similarity between the LiCoO$_2$ and Li$_2$O$_2$.}
    \label{fig:geo_structures_three}
\end{figure*}

We also investigated how the magnetic moments change with each round. Note that the GGA+U scheme which was employed in this study for DFT calculations adds artificial Coulomb interaction of localized electrons and often fails to predict the accurate value of magnetic moment, especially for transition-metal oxides~\cite{Zhou2006,Lee2017,REN2019362}. Hence, the magnetic moments obtained in this study will be used only to estimate a general tendency, not exact magnetic interactions between ions. Two inset figures in FIG.~\ref{fig:training_set_result}(b) show how magnetic moments of Co ions evolve as the rounds proceed. It is shown that the magnetic moment has values localized mostly around 0 or 1 at the 1st round while it is more dispersed after the 6$^{\text{th}}$ round. In particular, the population of 3$\mu_B$ bin grows significantly. According to the one-to-one comparison of magnetic moment between GGA+U and hybrid-DFT calculations (using the Heyd–Scuseria–Ernzerhof (HSE) functional~\cite{Heyd2003}), which provides more reliable value of magnetic moment, 3$\mu_B$ from GGA+U corresponds to ~0.11$\mu_B$ from HSE while the ones less than 3$\mu_B$ corresponds to zero~\cite{Lee2017}. The population of the data instances which have Co ions with greater than 3$\mu_B$ magnetic moment was 2.7\% and 18\% of the training set at the 1st and the 6th round, respectively.
This result implies that the magnetic moments of Co ions were predicted to be zero in most data instances in initial rounds but were corrected in later rounds by the anomaly detection and treatment process.
%
%
It also suggests to consider magnetic interactions between Co ions to better understand the thermodynamic behavior of the LCO system.

\begin{figure*}
    \centering
    \includegraphics[width=5in]{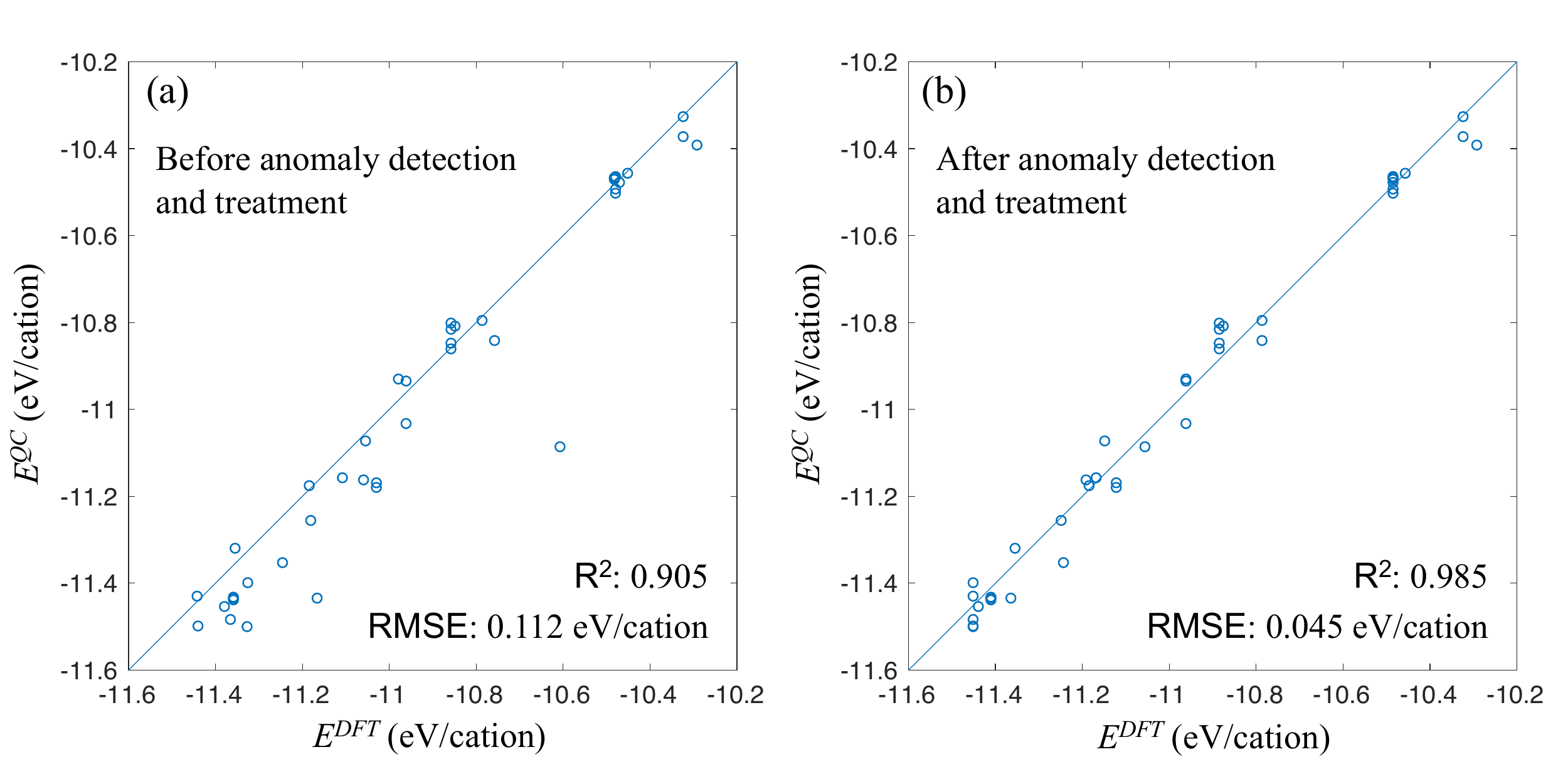}
    \caption{Comparison between $E_i^{DFT}$ and $E_i^{QC}$ for data instances in the test set of the 8 cationic lattice site LCO system -- (a) before and (b) after the anomaly detection and treatment were applied. RMSE and R$^2$ denote the resultant cost and the coefficient of determination, respectively.}
    \label{fig:test_set_result}
\end{figure*}

The optimized coefficients were then used to predict the energy for the test set (data instances that were not included to the training set), to investigate whether the optimization was over-fitted to the training set. As the incompatibility between the Co-rich region and the other regions was already indicated, all the data instances in the test set were for the ones with $2-x<1.5$. 40 data instances were prepared considering a typical ratio of the test set to the training set. The results are shown in FIG.~\ref{fig:test_set_result}. Although the cost was as high as 0.112 eV per cation without anomaly detection and treatment, it reduced to 0.045 eV per cation after the anomaly detection and treatment were applied. $R^2$ also increased from 0.905 to 0.985. The corresponding MAPE was 0.34\%.

\section{\label{sec:conclusion}Conclusion and Outlook}

In summary, we developed a heuristic quantum-classical algorithm to model the energy of substitutionally disordered binary crystalline materials as a function of atomic species on lattice sites via iterative learning based on the data of classically-computed quantum chemical simulation results but in quantum-enhanced feature space. We expect a quantum circuit model to be particularly suitable for representing data generated from a quantum mechanical system. The developed algorithm is expected to bring two computational advantages – the number of parameters increases linearly with the number of lattice sites with no truncation of interaction length-scale and anomalous data can be detected and treated efficiently as demonstrated on the LCO system. We remark that these advantages were possible because the energy of material systems is governed by only a few representative parameters while it is still computationally expensive to identify such parameters classically among all possible candidate parameters). Although it was demonstrated on relatively small size systems, the developed algorithm should be applicable to larger systems. 
While in this study the anomaly of data was assessed using the magnetic moments of ions, other types of material properties, such as geometric distortion can be used as well. We believe this study encourages follow-up experimental study with running the algorithm on quantum computing hardware. In this case, the entangling operators in the circuit can also be rearranged within the same block of entangling operators to satisfy conditions required for hardware experiments, giving us the flexibility of using a quantum circuit model that grows either linearly or with a constant depth as the system increases. 

Approaching the field of materials informatics with quantum circuit models opens up a lot of different avenues to explore. We leave the readers with some of the questions we find interesting for future work: how to efficiently tailor a quantum circuit model and measurement operators for a given lattice system, theoretical modelling and analysis of the computational cost (running time) of the quantum circuit model and its comparison with that of classical models as a function of growing system size, analyzing properties other than energies using quantum circuit models that are hard to generate with classical algorithms to name a few.


\section{\label{sec:Ackn}Acknowledgments}
The authors thank Julia Rice (IBM) and Mario Motta (IBM) for useful discussions, suggestions and help with figures. A part of this research used resources of the National Energy Research Scientific Computing Center (NERSC), a U.S. Department of Energy Office of Science User Facility operated under Contract No. DE-AC02-05CH11231.

\appendix
\section{\label{sec:DFT_calculations}DFT Calculations of the LCO System}
\begin{figure}[h]
    \centering
    \includegraphics[width=4.5in]{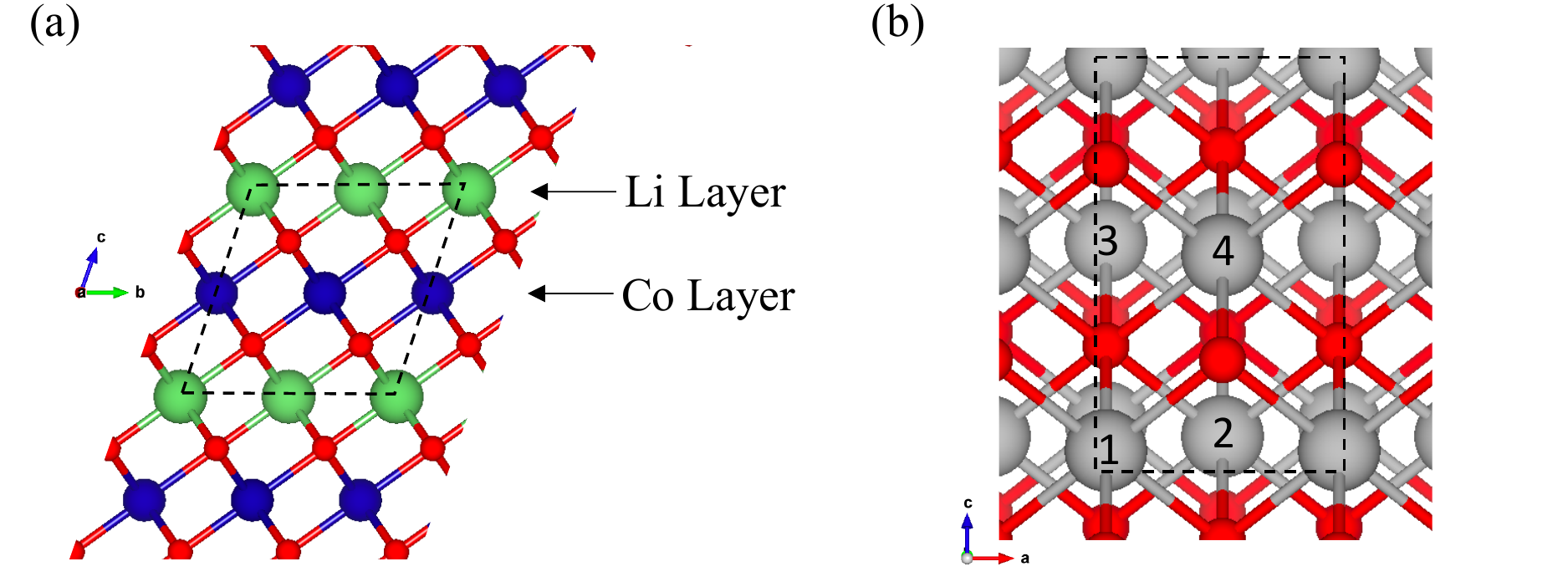}
    \caption{Illustration of (a) the ideal structure of layered Li-cobaltate and (b) 4 cationic lattice sites used to define the supercell of the 4 cationic lattice site system. The supercell is also indicated by the regions enclosed by dashed lines. In (b), the cationic sites are colored gray because they are assumed to be occupied by either Li or Co.}
    \label{fig:lco_host}
\end{figure}

The layered Li-cobaltate system has $R\bar{3}m$ space group with alternating layers of Li and Co ions, as illustrated in FIG.~{\ref{fig:lco_host}}. In this study, the same layered structure is adopted as the host structure of the LCO system but each lattice site is assumed to be occupied by either Li or Co ion without hard-separating Li and Co layers, considering the occurrence of cation-mixing. The supercell of the 4 cationic lattice site system was defined by taking the region illustrated in FIG.~\ref{fig:lco_host}(a).

The energies of the LCO system, $E^{DFT}$, were obtained classically with DFT using the generalized gradient approximation (GGA) with the Perdew-Burke-Ernzerhof parametrization~\cite{Perdew1997}, as implemented in Vienna Ab-initio Software Package (VASP)~\cite{Kresse1994,Kresse1993,Kresse1996,Kresse1996a}. The +U scheme is employed to account for the effect from electron localization~\cite{PhysRevB.57.1505,PhysRevB.52.R5467}, which is typically required in the DFT calculation for transition metal oxides. U value of 3.4 is chosen for Co ions. A cutoff energy of 520 eV is used and the k-point mesh is adjusted to ensure convergence of 1 meV per atom. The volume and shape of the supercell are allowed to change during the relaxation. While Co ions in LiCoO$_2$ (the LCO at $x$=1) are in general known to be Co$^{3+}$ and nonmagnetic, they may have different oxidation states and magnetic states at other compositions. Hence, the spin-polarized DFT calculations are performed in this study. An initial value of magnetic moment is given to each ion and it is allowed to change during relaxation. In principle, the magnetic moments are supposed to relax to the ones of the ground state regardless of the initial values. However, in practice, the relaxation of magnetic moments is very slow and several meta-stable states can be predicted for one identical $\sigma$~\cite{Zeller2006,Wills2010,Lee2017,Bihlmayer2018}, creating anomalous data. In this study, we assigned 0.05 for the initial values of magnetic moment. After the calculation is finished, the value of $\sigma_j$ is determined by the species of the atom located within a certain distance from the lattice site $j$. We used 0.35 {\AA} for that distance in this study; the distance between the nearest neighbor cations is around 2.85 {\AA} in the LCO system. 

Although $E^{DFT}$ values can be used directly in the optimization of the cost function given in Eq. (\ref{eq:cost_function}), in general the optimization becomes more efficient when the scale of variation in training values is smaller. For this purpose, $xE^0_{Li}+(1-x)E^0_{Co}$ is subtracted from $E^{DFT}$ values (eV per cation) and then used during the optimization of the circuit parameters. Note that $E^0_{Li}$ and $E^0_{Co}$ are not reference energies but artificially-designed factors to disperse the converted $\tilde{E}^{DFT}$ evenly in [-,+] range. In this study, $E^0_{Li}$ and $E^0_{Co}$ are set to -10.39 eV and -11.65 eV, respectively. Once the parameters are optimized, $xE^0_{Li}+(1-x)E^0_{Co}$ are added back to the values evaluated from the circuit to convert to $E^{QC}$ (eV per cation).

\section{\label{sec:anomaly_treatment}Anomaly Treatment}
Each data instance with large $\left|E^{QC}-E^{DFT}\right|$ based on $tol_{3}$ is addressed by performing additional DFT calculations and treated according to the following procedure.
\begin{enumerate}
    \item One DFT calculation is performed continuing from the previous DFT calculation results, the geometry of the supercell, the atoms' position, and the magnetic moments, to confirm that there were no numerical artifacts in the previous DFT calculation.
    \item DFT calculations are performed again with the magnetic moments initialized to values obtained from the other DFT calculations in the training set that satisfy the tolerance criteria and have a similar atomic configuration. If the energy returned is lower, the data point is updated in the training set, i.e. the old energy is replaced with the new energy.
    \item If lower energy is not predicted from the second step, DFT calculation is performed for a new $\sigma$ with a single lattice site, $\sigma_j$, altered. The altered lattice site is selected randomly. This new data instance is then added to the training set.
\end{enumerate}
The second step was essential for this study, because the LCO system contains transition-metal ions which we believe are affected significantly by magnetic interactions if magnetized. 
In the initial rounds, the model is probably still unconverged and the anomaly detection based on this model can be ineffective as well. In this case, referencing to the data instances qualified as being ‘correct’ (satisfying the tolerance criteria) will not be an efficient approach for selecting the initial values of magnetic moments in the follow-up DFT calculation. Thus, in the initial rounds we used a different approach instead of the one described above based on the considerations of the configuration of Li ions residing in the same layer as the Co ions. For example, if Co ions are surrounded by zero or negligible number of Li ions (for e.g. Co ions fully occupying one layer), those Co ions will be likely to have Co3+ oxidation state, which is non-magnetic. In this case, Co ions in the follow-up DFT calculation were assigned 0.05 for the initial magnetic moments if they were predicted to have non-zero (bigger than 0.05) magnetic moment in the previous DFT calculation. On the other hand, if Co ions are surrounded by Li ions densely, the oxidation states of those Co ions are probably different from 3+. In this case, the Co ions in the follow-up DFT calculation were assigned 1.05 and 3.05 if they were predicted to be non-magnetic in the previous DFT calculation. In the later rounds, the correlation between the distribution of Co ions and their magnetic moments tends to manifest as the model starts converging, and thus we can follow the second step as described above, i.e. adopting the values from the previous DFT calculations for similar configurations that satisfied the tolerance criteria for anomaly detection.
The second step may be less significant in the modeling of nonmagnetic materials. 

\section{\label{sec:qc_circuit_properties}Properties of the Quantum Circuit}
One-qubit and two-qubit operators in the proposed circuit are used to represent the interactions from atoms. For example, the interaction energy of a pair of atoms (equivalently a pair cluster) $\sigma_1\sigma_2$ can be explained directly by ENT($q_1$,$q_2$), which indicates the entanglement between qubit 1 and qubit 2, while the interaction energy of $\sigma_1\sigma_3$ cluster can be explained indirectly by a combination of entangling operations ENT($q_3$, ENT($q_1$,$q_2$)) and ENT($q_2$,$q_3$).  

As the occupation variables are encoded in $Z$-axis only, rotations along the $X$-axis, $R_X$ operators, are also added to provide more degrees of freedom. The inclusion of an additional layer of the single and two qubit operators provides more degrees of freedom and entangling operations to generate a richer feature space for describing the energy of the system. In principle, more layers could be added at the cost of more free parameters for optimization. The quantum circuit ansatz used is chosen to be hardware efficient for a system made of superconducting qubits with predominantly linear connectivity.

The choice of the operator $X_1Y_2X_3\cdots X_{N-1}Y_N$ for evaluating the energy of the system was made to overcome the inherent parity symmetries in different configurations since $\sigma_i$ takes values of $\pm$1 only. For example, operator with tensor product of all Z operators for $N$ qubits would not be able to distinguish between configurations having equal number of +1s. Another suitable choice for the measurement operators would be to randomly pick one of the three Pauli operators for each lattice site and keep it fixed throughout the implementation. Note that, while in this study we have chosen a single operator to evaluate the energy, a linear combination of multiple $N$ qubit operators can also be used but may require additional optimization parameters.

\begin{figure*}
    \centering
    \includegraphics[width=7in]{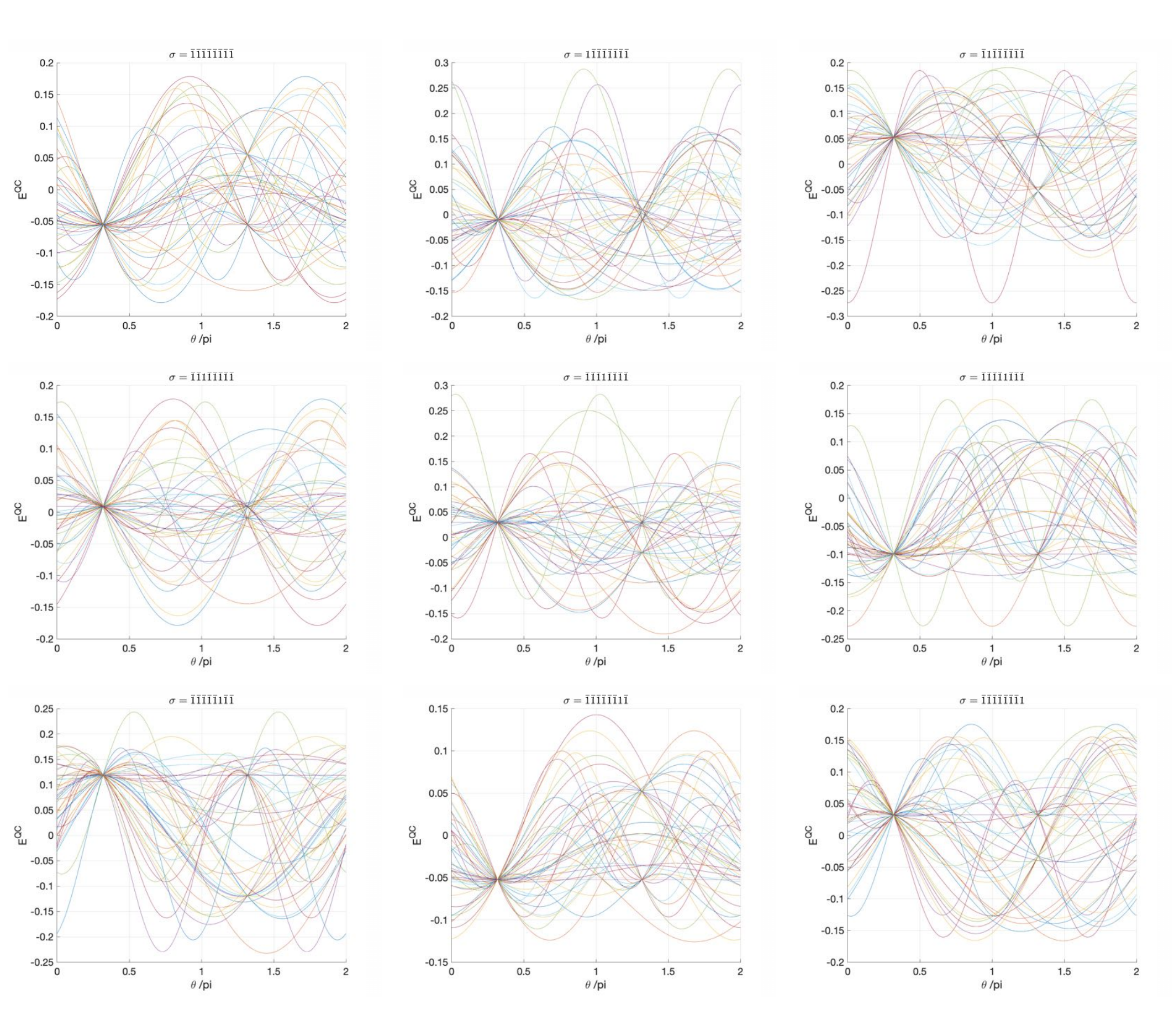}
    \caption{Variation of the individual parameters of the circuit for a sample of different 8 lattice site configurations. The energy of the system is calculated by fixing every parameter except for one. This is implemented for every parameter. Different colors in the above plots correspond to the free parameters of the circuit.}
    \label{fig:sensitivity_test}
\end{figure*}

The coefficients in the proposed circuit should be able to distinguish the different set of occupation variables effectively. This test is conducted by calculating $E^{QC}(\sigma)$ while varying $\sigma$ and the value of one coefficient in the range of $[0, 2\pi]$ (other coefficients are fixed to 1.0). If a coefficient produces a same constant outcome for all the $\sigma$'s, it will fail to distinguish between different $\sigma$. Note that the existence of a coefficient that results in constant energy is acceptable because it can be used to account for the reference energy, linear terms, and background errors. We examined 9 different $\sigma$'s: $\bar{1}\bar{1}\bar{1}\bar{1}\bar{1}\bar{1}\bar{1}\bar{1}$, $1\bar{1}\bar{1}\bar{1}\bar{1}\bar{1}\bar{1}\bar{1}$, $\bar{1}1\bar{1}\bar{1}\bar{1}\bar{1}\bar{1}\bar{1}$, $\cdots$, $\bar{1}\bar{1}\bar{1}\bar{1}\bar{1}\bar{1}\bar{1}1$ where $\bar{1}$ represents -1. The results are illustrated in FIG.~\ref{fig:sensitivity_test}. It is shown that each parameter produces different values of $E^{QC}$ for different value of occupation variables, which illustrates its capability to distinguish distinct $\sigma$ effectively. There are four constant coefficients which produce a constant $E^{QC}$ in the entire range $[0, 2\pi]$, however, we remark that those constants vary with $\sigma$'s.

\begin{figure*}
    \centering
    \includegraphics[width=6in]{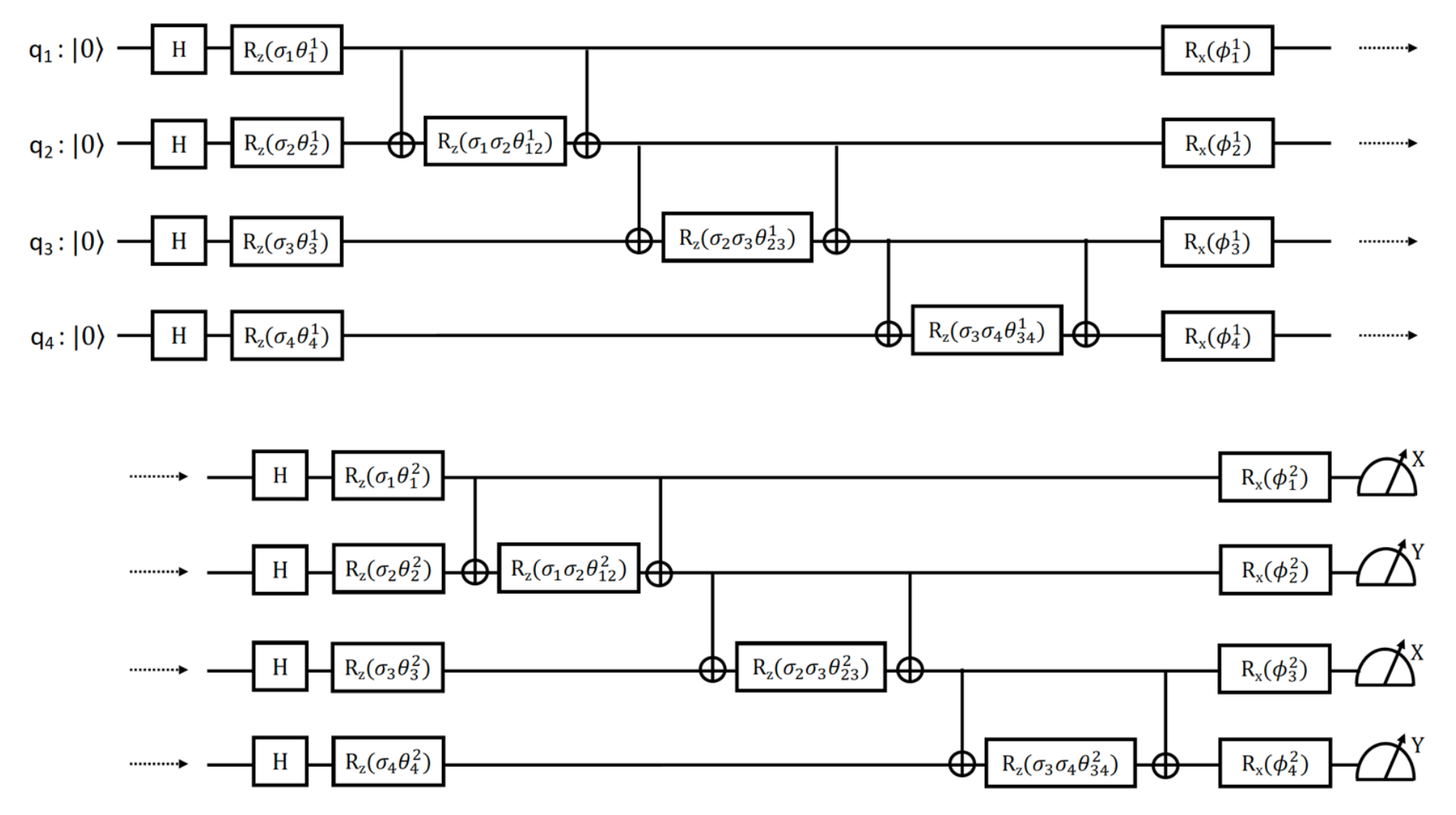}
    \caption{Detailed illustration of the quantum circuit used for a 4 lattice site system. $H$ represents Hadamard gates, $R_x$ and $R_z$ are rotations along the $X$ and the $Z$ axis respectively with the specified angles in brackets.}
    \label{fig:qc_exp_detail}
\end{figure*}

In FIG.~\ref{fig:qc_exp_detail}, we provide, as an example, the representation of the circuit used for the 4 lattice site system.  A similarly constructed 8 qubit circuit was used for the 8 lattice site system.



\newpage
\bibliography{qce}

\end{document}